\documentclass[aps,prc,twocolumn,showpacs,superscriptaddress,floatfix]{revtex4-1}  
\usepackage{graphicx}
\usepackage{amsmath}
\usepackage{multirow}

\newcommand {\nc} {\newcommand}
\nc {\beq} {\begin{eqnarray}} \nc {\eol} {\nonumber \\} \nc {\eeq}
{\end{eqnarray}} \nc {\eeqn} [1] {\label{#1} \end{eqnarray}} \nc
{\eoln} [1] {\label{#1} \\} \nc {\ve} [1] {\mbox{\boldmath $#1$}}
\nc {\rref} [1] {(\ref{#1})} \nc {\Eq} [1] {Eq.~(\ref{#1})} \nc
{\re} [1] {Ref.~\cite{#1}} \nc {\dem} {\mbox{$\frac{1}{2}$}} \nc
{\arrow} [2] {\mbox{$\mathop{\rightarrow}\limits_{#1 \rightarrow
#2}$}}
\begin{document}
\title{Astrophysical $S$ factor and reaction rate of the direct $^{12}{\rm C}(p, \gamma)^{13}{\rm N}$ capture process within a potential model approach
}

\author{E.M. Tursunov} \email{tursune@inp.uz}
\affiliation {Institute of Nuclear Physics, Academy of Sciences,
100214, Ulugbek, Tashkent, Uzbekistan} \affiliation {National
University of Uzbekistan, 100174 Tashkent, Uzbekistan}
\author{S.A. Turakulov}
\email{turakulov@inp.uz} \affiliation {Institute of Nuclear Physics,
Academy of Sciences, 100214, Ulugbek, Tashkent,
Uzbekistan}\affiliation {Tashkent State Agrarian University, 100140
Tashkent, Uzbekistan}
\author{A.S. Kadyrov}
\email{a.kadyrov@curtin.edu.au} \affiliation{Department of Physics
and Astronomy, Curtin University, GPO Box U1987, Perth, WA 6845,
Australia}

\begin{abstract}
The astrophysical direct nuclear capture reaction $^{12}{\rm C}(p,
\gamma)^{13}{\rm N}$ is studied within the framework of a potential
model. Parameters of the nuclear $p-^{12}$C interaction potentials
of the Woods-Saxon form are adjusted to reproduce experimental
$p-^{12}$C scattering phase shifts, as well as the binding energies
and empirical values of the asymptotic normalization coefficient
(ANC) for the $^{13}$N(1/2$^-$) ground state from the literature.
The reaction rates are found to be very sensitive to the description
of the value of the ANC of the $^{13}$N($1/2^{-}$) ground state and
width of the $^{13}$N($1/2^+$) resonance at the $E_x=2.365$ MeV
excitation energy. The potential model, which yields the ANC value
of 1.63 fm$^{-1/2}$ for the $^{13}$N($1/2^{-}$) ground state and a
value $\Gamma$=39 keV for the $^{13}$N($1/2^+$) resonance width, is
able to reproduce the astrophysical $S$ factor in the energy
interval up to 2 MeV, the empirical values of the reaction rates in
the temperature region up to $T=10^{10}$ K of the LUNA Collaboration
and the results of the R-matrix fit. The astrophysical factor
$S(0)=1.35$ keV b is found using the asymptotic expansion method of
D. Baye. The obtained value is in a good agreement with the Solar
Fusion II result. At the same time, the calculated value of 1.44 keV
b of the astrophysical $S$ factor at the Solar Gamow energy is
consistent with the result of the R-matrix fit of
$S(25~\rm{keV})=1.48 \pm 0.09$ keV b by Kettner {\it et al.}, but
slightly less than the result of $S(25~\rm{keV})=1.53 \pm 0.06$ keV
b the LUNA Collaboration.

\end{abstract}

 \maketitle

\section{Introduction}

\par Over the recent years, the direct nuclear capture reaction $^{12}{\rm C}(p, \gamma)^{13}{\rm N}$
 has been extensively studied both from theoretical and experimental
sides as the starting point of the CNO cycle in the hydrogen burning
process in stars, more massive than the Sun, especially in low mass
Asymptotic Giant Branch (AGB) and Red Giant Branch (RGB)
stars~\cite{il15,bor20}. This synthesis also has a strong influence
on the formation of the $^{12}$C/$^{13}$C abundances ratio and
$^{13}{\rm C}(\alpha, n)^{16}{\rm O}$ neutron source in the AGB
stars~\cite{clay2003}. Recently, updated Solar Fusion III review
\cite{achar24} highlighted the investigation of this reaction with
both the precise experimental techniques as well as various
theoretical models. In this regard, the application of empirical ANC
values, determined via indirect methods, plays a significant role
for the description of nuclear astrophysics processes in the stellar
environment.

\par The direct experimental study of the
$^{12}{\rm C}(p, \gamma)^{13}{\rm N}$ reaction is currently included
into the research plans of the leading experimental groups around
the world \cite{skow23,luna23,luna25,gyur23a,gyur23b,kett23}. The
first measurements of this reaction were performed 90 years ago
\cite{coc34,haf35}. In addition, a number of other important
experiments have been performed in the region around resonance
energy ($E_x$=2.365 MeV; $1/2^+$ and $E_x$=3.502 MeV; $3/2^-$)
\cite{bai50,hall50,heb60,young63,vogl63,rolfs74,bur08}. The most
important experimental data for the astrophysical $S$ factor has been
obtained by the LUNA (Laboratory for Underground Nuclear
Astrophysics) Collaboration down to the ultra-low  center of mass
(c.m.) energy of $E_{c.m.}$=68 keV thanks to the modern experimental
methods and techniques~\cite{luna23}. From the theoretical point of
view, the astrophysical $^{12}$C$(p,\gamma)^{13}$N direct capture
reaction has been studied wthin the framework of the
phenomenological R-matrix approach~\cite{azuma10,wies22}, potential
cluster models~\cite{huang10,irg18,irg20,dub25}, microscopic model
within the generator coordinate method~\cite{desc97}, and halo
effective field theory \cite{sad17}.

There are a number of works~\cite{fernan00,li10,art22,art02} devoted
to the extraction of an empirical ANC value of the virtual decay
$^{13}$N $\to$ $^{12}$C+$p$ of the $^{13}$N(1/2$^-$) ground state
from different proton transfer reactions. In particular, in
Ref.~\cite{fernan00} an empirical ANC value of $C_{1/2^-}=1.28\pm$
0.08 fm$^{-1/2}$ of the ANC has been extracted from the analysis of
the $^{12}\rm{C}(\rm{d,n})^{13}$N proton transfer reaction at the
laboratory energy of $E_{\rm{d}}$ = 12.4 MeV within the
distorted-wave Born approximation (DWBA). A different empirical ANC
value of $C_{1/2^-}=1.43\pm$ 0.06 fm$^{-1/2}$ was obtained within
the framework of the modified DWBA~\cite{art02} approach that takes
into account three-particle Coulomb effects by analyzing the
$^{12}\rm{C}(^3\rm{He},\rm{d})^{13}$N reaction at the energy
$E_{^3\rm{He}}$ = 22.3 MeV. A larger estimate of $C_{1/2^-}=1.64\pm$
0.11 fm$^{-1/2}$ was extracted from the analysis of the peripheral
nuclear reaction $^{12}\rm{C}(\rm{^7Li,^6He})^{13}$N  at
$E_{^7\rm{Li}}$ = 44.0 MeV energy within the DWBA~\cite{li10}.
Recent study within the framework of the combined analysis of the
$^{12}\rm{C}(^{10}\rm{B},^9\rm{Be})^{13}$N transfer reaction at the
energy $E_{^{10}\rm{B}}$ = 41.3 MeV using the modified DWBA and the
FRESCO code~\cite{art22} yielded an empirical ANC value of
$C_{1/2^-}=1.63\pm$ 0.13 fm$^{-1/2}$. The calculated value of ANC in
the source term approach using shell model wave functions
\cite{tim2013} of $C_{1/2^-}$=1.38 fm$^{-1/2}$ lies within the range
of the empirically extracted values.

Since for every bound state of the nucleus the ANC value is unique,
it is natural to ask a question, as to which of the above-mentioned
empirical ANC values, ranging from 1.28 fm$^{-1/2}$ to 1.64
fm$^{-1/2}$, is most realistic. This question is very important for
nuclear astrophysics, since the values of the astrophysical $S$
factor of the direct capture reaction $^{12}{\rm C}(p,
\gamma)^{13}{\rm N}$ at low astrophysically relevant energies are
mostly defined by this parameter \cite{mukhblokh}. In this respect
it is useful to develop potential models which can help to find the
most realistic ANC value with the help of the appropriate choice of
the potential parameters for the bound state in the description of
the existing experimental data for the astrophysical $S$ factor of
the LUNA Collaboration \cite{kett23,gyur23a,gyur23b,skow23,luna23}
at the lowest energy region. On the other hand, the potential model
can help to optimize the description of the astrophysical $S$ factor
and the reaction rate of the direct capture process in the region
including $1/2^+$ resonance energy of $E$=0.424 MeV by choosing the
most realistic $S$-wave potential parameters.

The aim of the present work is to construct the most realistic
potential model for the description of the existing experimental
data for the astrophysical $S$ factor and reaction rates of the
$^{12}{\rm C}(p, \gamma)^{13}{\rm N}$ direct capture process taking
into consideration the available empirical ANC values from the
literature~\cite{fernan00,li10,art22,art02}.

It was shown \cite{tur21,tur21b,tur23a,tur24} that the proposed
potential model can reproduce the empirical ANC value and bound
state energy of the chosen nucleus, independent of the form of
nuclear potential model used, whether it is the Woods-Saxon form or
the Gaussian one. Below, the Woods-Saxon potential parameters in the
partial $^2S_{1/2}$, $^2P_{1/2}$, $^2P_{3/2}$, $^2D_{3/2}$ and
$^2D_{5/2}$ initial scattering states are adjusted to reproduce the
experimental phase shifts \cite{dub08,brown67}. In this work, the
ability of the proposed potential model to describe the experimental
astrophysical $S$ factor and the reaction rates of the $^{12}{\rm
C}(p,\gamma)^{13}{\rm N}$ direct capture process will be examined.

The structure of the paper is as follows. The theoretical model is
briefly described in Section II. The numerical results are presented
in Section III and the conclusions are given in the last section.

\section{Wave functions and interaction potentials}

In this section, a general description of the model and the wave
functions are given. The two-body interaction potentials are also
presented. Details of the formalism used in our study have been
given in Refs.~\cite{tur21,tur21b,tur23a,tur24}. Within the
single-channel approximation, the wave functions of the initial
scattering and final bound states are presented as
\begin{eqnarray}
\Psi_{lS}^{J}=\frac{u_E^{(lSJ)}(r)}{r}\left\{Y_{l}(\hat{r})\otimes\chi_{S}(\xi)
\right\}_{J M}
\end{eqnarray}
and
\begin{eqnarray}
\Psi_{l_f S'}^{J_f}
=\frac{u^{(l_fS'J_f)}(r)}{r}\left\{Y_{l_f}(\hat{r})\otimes\chi_{S'}(\xi)
\right\}_{J_f M_f},
\end{eqnarray}
respectively.

The two-body Schr\"{o}dinger equation
\begin{align}
\left[-\frac{\hbar^2}{2\mu}\left(\frac{d^2}{dr^2}-\frac{l(l+1)}{r^2}\right)+V^{
lSJ}(r)\right] u_E^{(lSJ)}(r)= E u_E^{(lSJ)}(r),
\end{align}
is solved by using the Numerov algorithm of high precision.  The
radial parts of the initial scattering wave functions in the
$^2S_{1/2}$, $^2P_{1/2}$, $^2P_{3/2}$, $^2D_{3/2}$, $^2D_{5/2}$
partial waves of the $p-^{12}${\rm C} system are calculated
numerically. Hereafter, everywhere $V^{lSJ}(r)$ is a two-body
potential in the partial wave with orbital angular momentum $l$,
spin $S$ and total angular momentum $J$. The bound-state solution of
the  Schr\"{o}dinger equation yields a wave function
$u^{(l_fS'J_f)}(r)$ of the final $^2P_{1/2}$ ground state.

The central p-$^{12}${\rm C} two-body potential is chosen in the
Woods-Saxon form \cite{huang10,tur24}:
\begin{eqnarray}\label{pot}
V^{lSJ}(r)=V_{0}\left[1+\exp\left(\frac{r-R_0}{a_0}\right)\right]^{-1}+V_{c}(r),
\end{eqnarray}
where $V_0$ is depth of the central part of the potential, $R_0=r_0
\,A^{1/3}$ fm and $a_0$ are geometric parameters of the potential,
characterizing radius and diffuseness, respectively. Here, $A_1$ and
$A_2$ are the atomic mass numbers of the first (proton) and second
($^{12}${\rm C} nucleus) clusters, respectively, and $A=A_1+A_2$.
The Coulomb potential is given by the spherical charge
distribution~\cite{huang10,tur24}
\begin{eqnarray}\label{Coulomb}
 V_c(r)=
\left\{
\begin{array}{lc}
 Z_1 Z_2 e^2/r &  {\rm if} \,\, r>R_c, \\
Z_1 Z_2 e^2 \left(3-{r^2}/{R_c^2}\right)/(2R_c) & {\rm if} \,\,r \le
R_c,
\end{array}
\right.
\end{eqnarray}
with the Coulomb radius $R_c=1.25 A^{1/3}$ fm, and charge numbers
$Z_1$, $Z_2$ of the first and second clusters, respectively.

The Schr\"{o}dinger equation is solved numerically in the entrance
and exit channels with the two-body $p-^{12}\rm{C}$ nuclear
potentials of the Woods-Saxon form (\ref{pot}) with the
corresponding Coulomb potential of the spherical charge
distribution. Hereafter, everywhere  in the numerical calculations
the parameter values in the atomic mass units $\hbar^2/2 u$ =20.9008
MeV fm$^2$, 1u=931.494 MeV, $m_p =A_1$u=1.007276467u ,
m($^{12}$C)=$A_2$u=12u and $\hbar$c=197.327 MeV fm are used.

\begin{table*}[htbp]
\centering \caption{Parameters of the Woods-Saxon $p-^{12}$C
interaction potential models.} \label{tab1} {\footnotesize
\begin{tabular}{c c c c c c c c c} \hline \hline & $^{2S+1}L_J$ & $V_{0}$
(MeV)  & $r_0$ (fm) & $a_0$ (fm) & ANC (fm$^{-1/2})$ & ANC(exp.)
(fm$^{-1/2})$ &$E_{FS}^{^{13}\rm{N}}$ (MeV)\\\hline
\multirow{2}{3em}{$\textrm{V}_\textrm{M1}$} & $^2S_{1/2}$ & $-$29.87  & 1.90 & 0.14 &  & &$-$18.94 \\
& $^2P_{1/2}$ & $-$43.3413  & 1.14 & 0.44 & 1.63 &1.63$\pm$0.13 \cite{art22} &$-$\\
\hline
\multirow{2}{3em}{$\textrm{V}_\textrm{M2}$} & $^2S_{1/2}$ & $-$36.57  & 1.70 & 0.20 &  & &$-$23.24 \\
& $^2P_{1/2}$ & $-$48.4922  & 1.06 & 0.36 & 1.43 &1.43$\pm$0.06 \cite{art02} &$-$\\
\hline
\multirow{2}{3em}{$\textrm{V}_\textrm{M3}$} & $^2S_{1/2}$ & $-$45.37  & 1.50 & 0.30 &  & &$-$28.17 \\
& $^2P_{1/2}$ & $-$56.1195  & 0.97 & 0.32 & 1.28 & 1.28$\pm$0.08
\cite{fernan00}&$-$
\\\hline
\multirow{2}{3em}{$\textrm{V}_\textrm{M4}$} & $^2S_{1/2}$ & $-$29.84  & 1.90 & 0.20 &  & &$-$18.75 \\
& $^2P_{1/2}$ & $-$43.3413  & 1.14 & 0.44 & 1.63 &1.63$\pm$0.13 \cite{art22} &$-$\\
\hline
& $^2P_{3/2}$ & $-$728.67  & 0.22 & 0.19 & & &$-$\\
& $^2D_{3/2}$ & $-$24.56   & 0.98 & 0.82 & & &$-$\\
& $^2D_{5/2}$ & $-$69.36  & 1.15 & 0.55 & & &$-$\\
\hline \hline
\end{tabular}}
\end{table*}

\begin{figure}[htbp]
\includegraphics[width=\columnwidth]{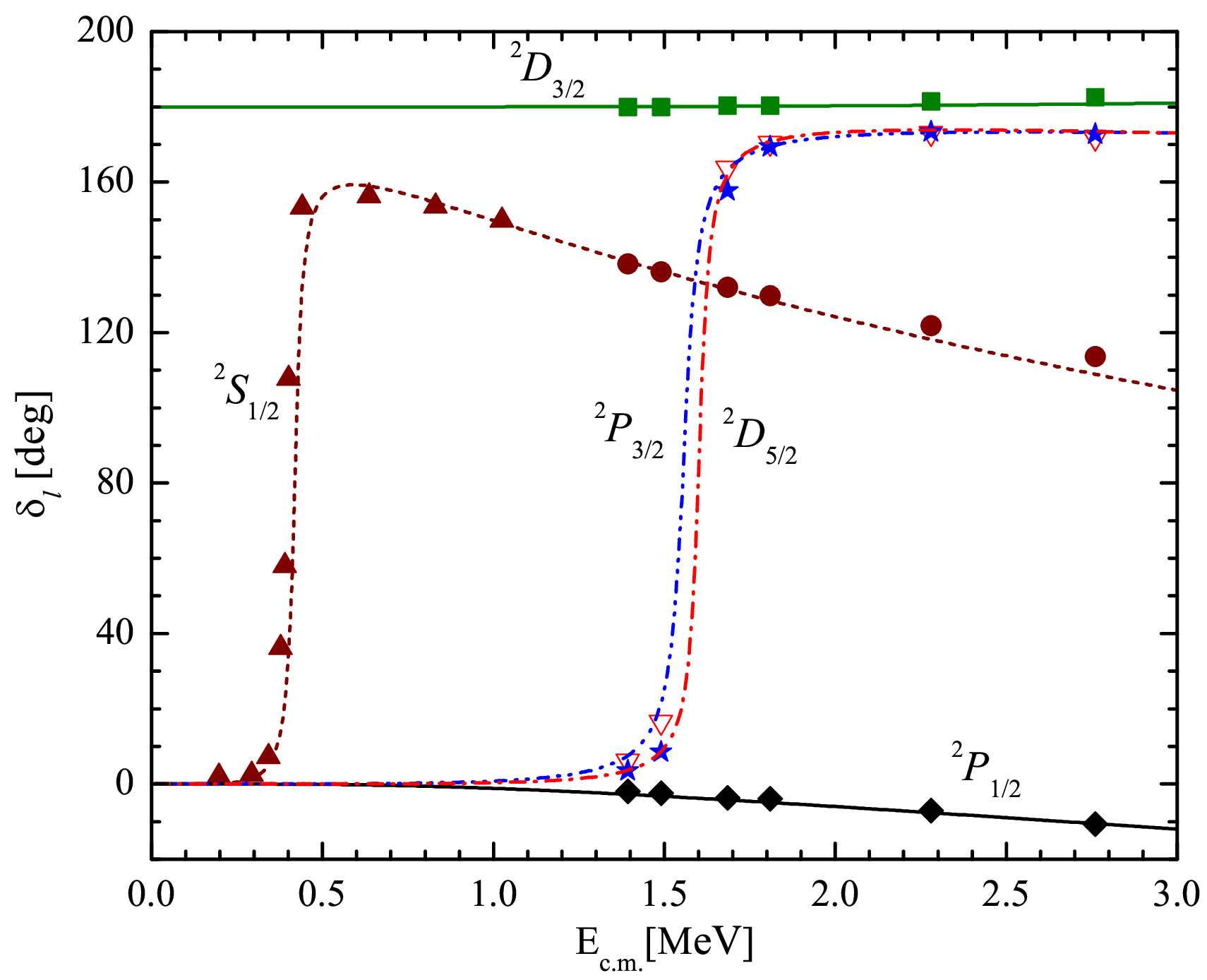} \caption{The phase shifts for $p-^{12}${\rm C} elastic scattering  within the
$\textrm{V}_\textrm{M1}$ potential model. The experimental data are
taken from Refs.~\cite{dub08,brown67}.} \label{fig1}
\end{figure}

The geometric parameters of the Woods-Saxon potential for different
versions of the model are given in Table \ref{tab1}. The proposed
potential models $\textrm{V}_\textrm{M${\rm{1}}$}$,
$\textrm{V}_\textrm{M${\rm{2}}$}$, $\textrm{V}_\textrm{M${\rm{3}}$}$
reproduce the energy $E(1/2^-)=-$1.9435 MeV and the ANC values 1.63
fm$^{-1/2}$ \cite{art22}, 1.43 fm$^{-1/2}$ \cite{art02} and 1.28
fm$^{-1/2}$ \cite{fernan00}, respectively of the $^{13}$N(1/2$^-$)
nucleus ground state which appears in the $^2P_{1/2}$ partial wave
of the $p+^{12}C$ system. The potentials in the $^2S_{1/2}$,
$^2P_{3/2}$, $^2D_{3/2}$ and $^2D_{5/2}$ partial waves are adjusted
to reproduce the experimental phase shift data. The theoretical
phase shift results are given in Fig.~\ref{fig1} in comparison with
the experimental data \cite{dub08,brown67}. The parameters of the
potential in the scattering states $^2P_{3/2}$, $^2D_{3/2}$ and
$^2D_{5/2}$ are common for all the models (see Table \ref{tab1}).

Model $\textrm{V}_\textrm{M${\rm{4}}$}$ differs from the model
$\textrm{V}_\textrm{M${\rm{1}}$}$ only in the $^2S_{1/2}$ partial
scattering wave. As was given in Table \ref{tab2}, the potential
models $\textrm{V}_\textrm{M${\rm{1}}$}$,
$\textrm{V}_\textrm{M${\rm{2}}$}$, and
$\textrm{V}_\textrm{M${\rm{3}}$}$ reproduce the experimental values
of the $^2S_{1/2}$ ($E_x=$2.365 MeV) resonance width
$\Gamma_{c.m.}$(exp.)=35.2$\pm$0.5 keV  and its energy position
$E_{res}=$422 keV \cite{gyur23b}. The only model
$\textrm{V}_\textrm{M${\rm{4}}$}$ yields slightly larger
$\Gamma_{c.m.}$=39 keV estimate for the $^2S_{1/2}$ resonance width.
The last model was introduced in order to examine the sensitivity of
the astrophysical $S$ factor and the reaction rates on the
description of the resonance width in the most important $^2S_{1/2}$
scattering state. As will be shown below, the reaction rates are
very sensitive to the description of the $^2S_{1/2}$ resonance
width.

 The empirical ANC values reported in Table \ref{tab1}
were derived from the analysis of the cross-section of the proton
transfer reaction using the modified DWBA \cite{art22,art02} and the
continuum discretized coupled channels (CDCC) method
\cite{fernan00}. The elastic $p-^{12}$C scattering in low energy
region was previously studied within the halo effective field theory
\cite{in24} and the multilevel, multichannel R-matrix approach
\cite{azuma10}. As can be seen from Fig.~\ref{fig1}, the
experimental phase shifts are well described within the proposed
potential models.

\begin{table*}[htbp]
\centering \caption {Parameters of the $S$-, $P$- and $D$- wave
resonances of the $^{13}$N nucleus, calculated within different
potential models, in comparison with the experimental data}
\label{tab2} {\footnotesize
\begin{tabular}{c c c c c c c c c c} \hline \hline $^{2S+1}L_J$ & $E_x$ & $J_i^{\pi}$
(MeV)  & $\Gamma_{c.m.}$(exp.) (keV) & $E_{res}$ (exp.) (keV) &
$\Gamma_{c.m.}$(theory) (keV) & $E_{res}$ (theory) (keV) \\\hline 
$^2S_{1/2}$ ($\textrm{V}_\textrm{M1}$) & 2.365 & $1/2^+$& 35.2$\pm$0.5 \cite{gyur23b} & 424.2$\pm$0.7 \cite{gyur23b} & 36  &422 \\

$^2S_{1/2}$($\textrm{V}_\textrm{M2}$) & &  &  &  & 35  &424 \\

$^2S_{1/2}$ ($\textrm{V}_\textrm{M3}$) & &  &  & & 35  &427 \\

$^2S_{1/2}$($\textrm{V}_\textrm{M4}$) & &  &  & & 39  &424 \\

$^2P_{3/2}$ & 3.502 & $3/2^-$& 62$\pm$4 \cite{ajzen91} & 1558.5$\pm$2 \cite{ajzen91} & 62  &1558\\
$^2D_{5/2}$ & 3.547 & $5/2^+$ & 45.2$\pm$0.5 \cite{kett23} & 1601.0$\pm$0.5 \cite{kett23} & 44  &1603\\
\hline \hline
\end{tabular}}
\end{table*}

The $p-^{12}$C system in the $S$-wave contains a Pauli forbidden
state according to the classification of orbital states of the
Young's diagrams \cite{dub25}. In the last column of Table
\ref{tab1}, the energy values of the Pauli forbidden state in the
$^2S_{1/2}$ partial wave are given for the different potential
models. Properties of the most important $^{13}$N$(1/2^+)$ resonance
($E_x$=2.365 MeV) in this partial wave have been studied within the
framework of different theoretical approaches in
Refs.~\cite{kett23,gyur23b,dub25,ajzen91,tur23b} and the width of
the resonance was estimated to belong to the 30.1 keV $-$ 37.8 keV
interval.


The resonance  energies $E_{res}$ and widths $\Gamma$ of the
$^{13}${\rm N} nucleus in the $^2P_{3/2}$ and $^2D_{5/2}$ partial
waves are also presented in Table ~\ref{tab2}. As can be seen from
the table, the existing experimental data
~\cite{kett23,gyur23b,ajzen91} are well described within the
developed potential model approach.


\section{Astrophysical $S$ factor of the $^{12}{\rm C}(p,
\gamma)^{13}{\rm N}$ direct capture reaction}

\begin{figure}[htbp]
\includegraphics[width=\columnwidth]{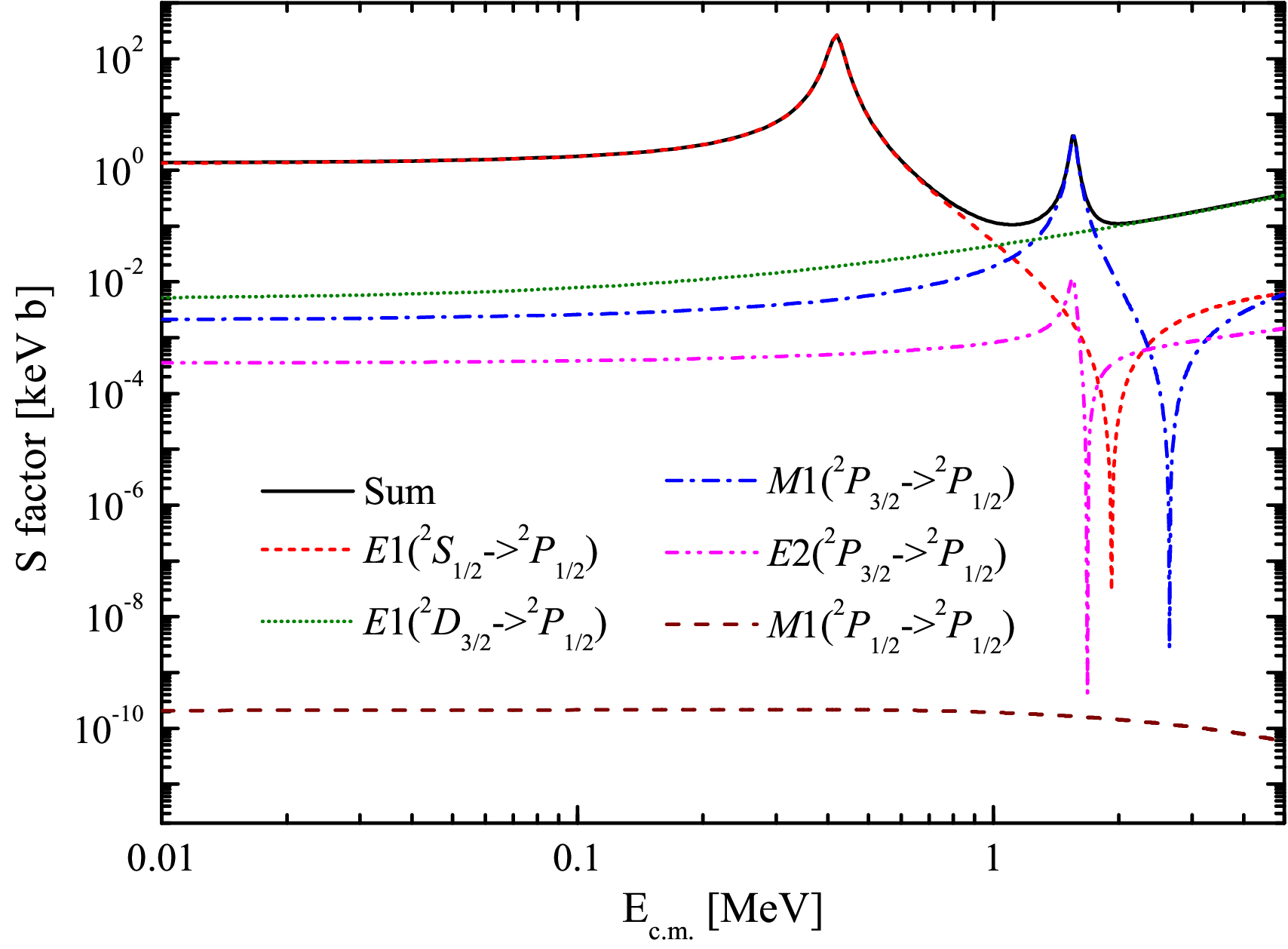}\caption{The $E$1, $E$2 and $M1$ partial
transition contributions to the astrophysical $S$ factor from
different initial scattering states to the final
$^{13}\rm{N(1/2^-)}$ ground state within the potential model
$\textrm{V}_\textrm{M1}$.} \label{fig2}
\end{figure}

The contributions of the electromagnetic $E1$, $E2$ and $M1$
transition operators into the astrophysical $S$ factor of the
$^{12}{\rm C}(p, \gamma)^{13}{\rm N}$  reaction are shown in
Figure~\ref{fig2} for the model $\textrm{V}_\textrm{M${\rm{1}}$}$
from Table \ref{tab1}. As is known from the literature \cite{bur08}
and as can be seen in the figure, the two resonant states $1/2^+$
($E_x$=2.365 MeV) and $3/2^-$ ($E_x$=3.502 MeV) yield the most
important contribution to the direct radiative capture process
through the electric $^2S_{1/2} \rightarrow ^2P_{1/2}$ and magnetic
$^2P_{3/2} \rightarrow ^2 P_{1/2}$ dipole transitions to the
$^{13}{\rm N}$ ground state. The transition from the $1/2^+$
resonance state completely dominates in the region below 1 MeV,
while the magnetic transition from the $^2P_{3/2}$ state dominates
beyond this region up to the reaction energy of about 2 MeV. This is
why the problem of the consistent description of the properties of
these resonances is of crucial importance.

As can be seen from Figure~\ref{fig2}, the electric $^2D_{3/2}
\rightarrow ^2 P_{1/2}$ dipole transition plays a minor role for the
process below 1 MeV. The contributions of the electric quadrupole
transition ($E$2) $^2P_{3/2}\rightarrow ^2 P_{1/2}$ is even more
suppressed, while the contribution from magnetic dipole ($M$1)
$^2P_{1/2}\rightarrow ^2 P_{1/2}$ transition operators is entirely
negligible in the astrophysical energy region.

\begin{figure}[htbp]
\includegraphics[width=\columnwidth]{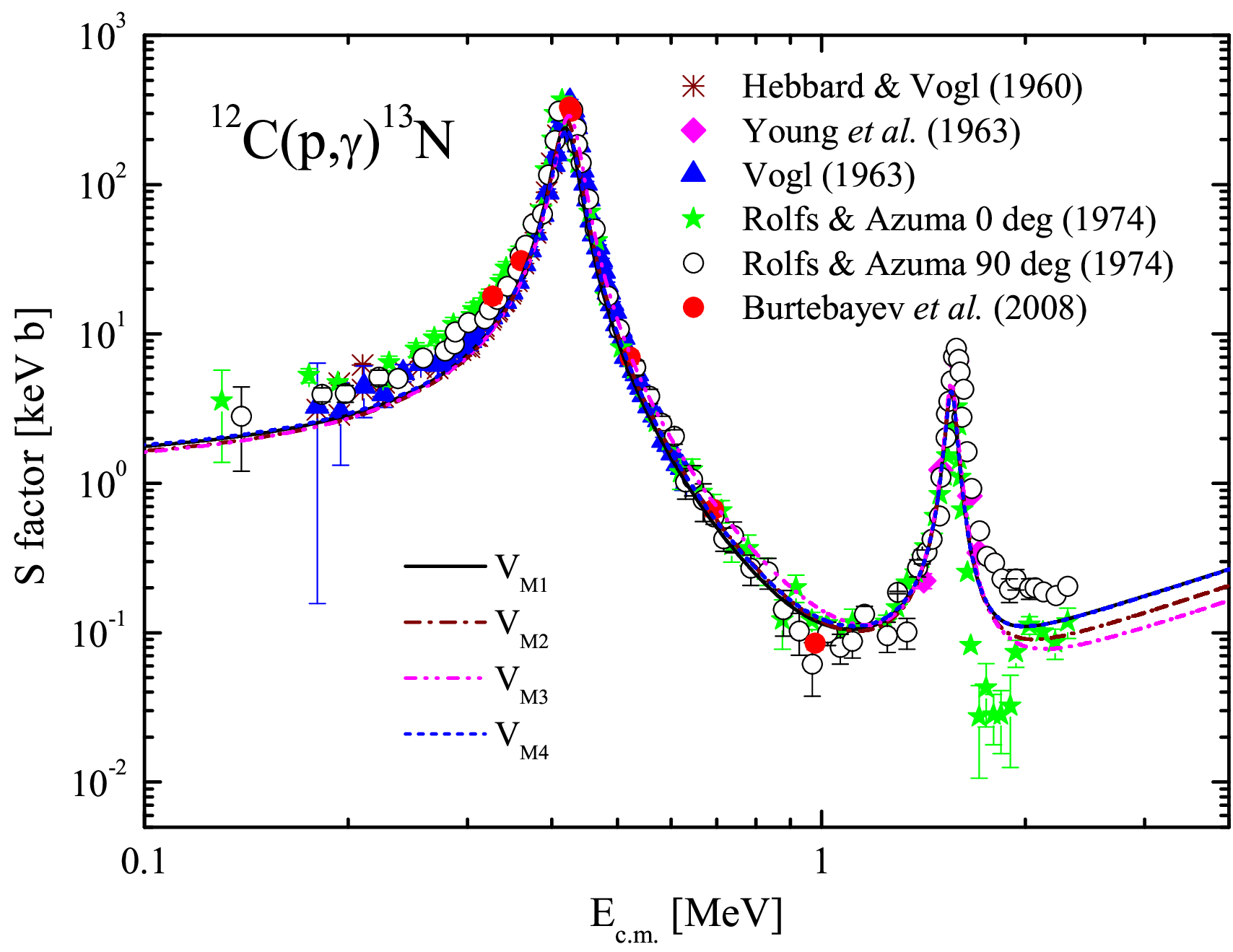}\caption{Astrophysical $S$ factor for the $^{12}$C(p,$\gamma)^{13}$N synthesis
reaction, estimated within different potential models in comparison
with the experimental data from Refs.
\cite{heb60,young63,vogl63,rolfs74,bur08}.} \label{fig3}
\end{figure}

\begin{figure}[htbp]
\includegraphics[width=\columnwidth]{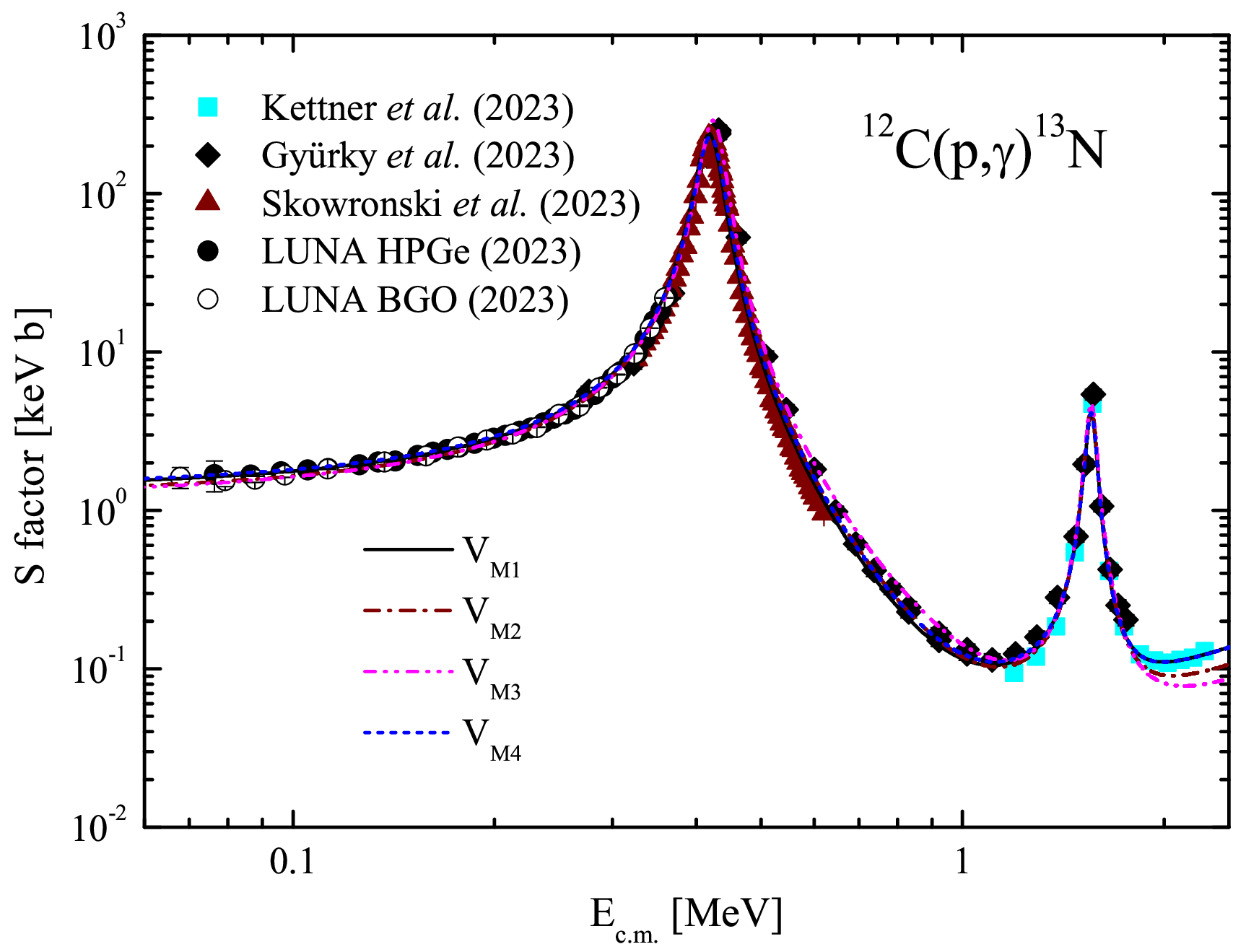}\caption{Astrophysical $S$ factor for the $^{12}$C(p,$\gamma)^{13}$N synthesis
reaction, estimated within different potential models in comparison
with the newest experimental data from Refs.
\cite{skow23,luna23,gyur23a,kett23}.} \label{fig4}
\end{figure}

In Fig.~\ref{fig3} the calculated astrophysical $S$ factors of the
direct radiative capture reaction $^{12}{\rm C}(p, \gamma)^{13}{\rm
N}$ within the potential models $\textrm{V}_\textrm{M${\rm{1}}$}$,
$\textrm{V}_\textrm{M${\rm{2}}$}$, $\textrm{V}_\textrm{M${\rm{3}}$}$
and $\textrm{V}_\textrm{M${\rm{4}}$}$ are compared with the old
experimental data sets from Refs.~
\cite{heb60,young63,vogl63,rolfs74,bur08}. And in Fig.~\ref{fig4},
the theoretical results are compared with the most recent
experimental data sets of the LUNA Collaboration ~\cite{luna23}
resulting from the direct measurements and other recent data sets
from Refs.~ \cite{skow23,gyur23a,kett23} within the whole
astrophysical low-energy region. As can be seen from the figures,
the agreement between theoretical results and experimental data sets
is very good. Clearly, the theoretical models describe the newest
data sets from 2023 more accurately. However, from these figures one
can not make a conclusion about the advantages or disadvantages  of
the above models since they all describe the existing data sets at
the same high level.

It should be noted that the old experimental data points were
obtained down to the center-of-mass energy of 130
keV~\cite{rolfs74}, while the newest experimental points shown in
Fig.~\ref{fig4} were determined down to extremely low energy of 68
keV~\cite{luna23} thanks to the condition in a background-shielded
at a depth of 1400 meters. However, the new experimental
points~\cite{skow23,gyur23a}, which correspond to the energy region
around the first resonance ($E_x$=2.365 MeV; $1/2^+$) peak, are
about 40-50 percent lower than the old
data~\cite{vogl63,rolfs74,bur08}. From this point of view, one can
argue that the best description of the new experimental data for the
astrophysical $S$ factor of the $^{12}$C(p,$\gamma)^{13}$N direct
capture process corresponds to the $\textrm{V}_\textrm{M1}$ and
$\textrm{V}_\textrm{M4}$ potential models. The latter yield the ANC
value of 1.63$\pm$0.13 fm$^{-1/2}$ for the $^{13}\rm{N(1/2^-)}$
ground state \cite{art22}.

In Table ~\ref{tab3} the calculated values of the astrophysical
$S(E)$ factor of the direct $^{12}$C(p,$\gamma)^{13}$N capture
process at zero and at the solar Gamow peak $S(25 ~\rm{keV})$ energy
are given for the potential models developed above. The zero-energy
astrophysical $S(0)$ factor was estimated by using the asymptotic
expansion method \cite{baye00}. The method was previously used to
calculate $S(0)$ factors for the $^3$He$(\alpha,\gamma)^7$Be,
$^3$H$(\alpha,\gamma)^7$Li, and $^7$Be$(p,\gamma)^8$B capture
reactions resulting estimates consistent with other available data
\cite{tur23c}. The theoretical results obtained within the all four
potential models developed in present work, particularly are
consistent with the Solar Fusion II result  of $S(0)=1.34 \pm 0.21$
keV b \cite{adel11}.

\begin{table}[htbp]
\centering \caption{The calculated values of the astrophysical
$S(E)$ factor at the zero and solar Gamow energies.}
 \begin{tabular}{c c c c c} \hline \hline
E$^{\rm{*}}~\rm{(keV)}$ & \multicolumn{4}{c}{ S(E)~\rm{(keV b)}}\\
  & $\rm{V}_{\rm{M1}}$   & $\rm{V}_{\rm{M2}}$   & $\rm{V}_{\rm{M3}}$ & $\rm{V}_{\rm{M4}}$ \\
\hline
0.0& 1.33 & 1.21 &1.17 &1.35\\
25 & 1.41 & 1.29 &1.25 &1.44\\
\hline \hline
\end{tabular} \label{tab3}
\end{table}

On the other hand, the calculated values of the astrophysical $S$ 
factor at the solar Gamow energy 1.41 keV b and 1.44 keV b within
the $\textrm{V}_\textrm{M${\rm{1}}$}$ and
$\textrm{V}_\textrm{M${\rm{4}}$}$ models, respectively, are in good
agreement with the result of the R-matrix fit of
$S(25~\rm{keV})=1.48 \pm 0.09$ keV b  by Kettner {\it et al.}
\cite{kett23}, but slightly lower than the result of
$S(25~\rm{keV})=1.53 \pm 0.06$ keV b by the LUNA Collaboration
\cite{luna23,achar24}. At the same time, the models
$\textrm{V}_\textrm{M${\rm{2}}$}$ and
$\textrm{V}_\textrm{M${\rm{3}}$}$ yield somewhat smaller estimates
for this value.

\section{Reaction rates of the $^{12}{\rm C}(p, \gamma)^{13}{\rm
N}$ astrophysical direct capture process}

For the estimation of chemical element abundances in the
stars, one needs to evaluate the values of nuclear reaction rates on
the basis of the calculated cross-sections. This quantity cannot be
extracted from the data of the experimental measurements. The
well-known expression for the reaction rate $N_{A}(\sigma v)$
 reads \cite{nacre,fow1975}
\begin{eqnarray}
N_{A}(\sigma v)=N_{A}
\frac{(8/\pi)^{1/2}}{\mu^{1/2}(k_{\text{B}}T)^{3/2}}
\int^{\infty}_{0} \sigma(E) E \exp(-E/k_{\text{B}}T) d E,
\end{eqnarray}
where $\sigma(E)$ is the calculated cross-section of the process,
 $k_{\text{B}}$ is the Boltzmann coefficient, $T$ is the
temperature, $N_{A}=6.0221\times10^{23}\, \text{mol}^{-1}$ is the
Avogadro number.

\begin{table*}[htbp]
\centering \caption{The $^{12}{\rm C}(p, \gamma)^{13}{\rm N}$
astrophysical reaction rates $N_{A}(\sigma v)$ ($ \textrm{cm}^{3}
\rm{mol}^{-1} \rm{s}^{-1}$) in the temperature interval $ 0.001\leq
T_{9} \leq 10 $ calculated within the $\textrm{V}_\textrm{M1}$ and
$\textrm{V}_\textrm{M4}$ potential models.}
\begin{tabular}{c c c c c c c c} \hline \hline
~~~$T_{{\rm{9}}}$ ~~&  $\textrm{V}_\textrm{M1}$~~~~~&~~
$\textrm{V}_\textrm{M4}$~~~~~&~~~
$T_{{\rm{9}}}$~~~ &~~$\textrm{V}_\textrm{M1}$&$\textrm{V}_\textrm{M4}$~~~~~&\\
\hline
0.001&$6.84\times 10^{-51}$ &$6.99\times 10^{-51}$&0.14&$3.74\times 10^{-4}$&$3.83\times 10^{-4}$\\
0.002&$7.98\times 10^{-39}$ &$8.16\times 10^{-39}$&0.15&$6.68\times 10^{-4}$&$6.84\times 10^{-4}$\\
0.003&$5.66\times 10^{-33}$ &$5.79\times 10^{-33}$&0.16&$1.14\times 10^{-3}$&$1.17\times 10^{-3}$\\
0.004&$2.76\times 10^{-29}$ &$2.82\times 10^{-29}$&0.18&$2.95\times 10^{-3}$&$3.03\times 10^{-3}$\\
0.005&$1.16\times 10^{-26}$ &$1.18\times 10^{-26}$&0.20&$6.81\times 10^{-3}$&$6.97\times 10^{-3}$\\
0.006&$1.15\times 10^{-24}$ &$1.18\times 10^{-24}$&0.25&$4.01\times 10^{-2}$&$4.09\times 10^{-2}$\\
0.007&$4.55\times 10^{-23}$ &$4.66\times 10^{-23}$&0.30&$1.89\times 10^{-1}$&$1.90\times 10^{-1}$\\
0.008&$9.42\times 10^{-22}$ &$9.65\times 10^{-22}$&0.35&$7.60\times 10^{-1}$&$7.57\times 10^{-1}$\\
0.009&$1.22\times 10^{-20}$ &$1.25\times 10^{-20}$&0.40&$2.50\times 10^{0}$&$2.48\times 10^{0}$\\
0.010&$1.11\times 10^{-19}$ &$1.13\times 10^{-19}$&0.45&$6.74\times 10^{0}$&$6.66\times 10^{01}$\\
0.011&$7.60\times 10^{-19}$ &$7.78\times 10^{-19}$&0.5&$1.52\times 10^{1}$&$1.51\times 10^{1}$\\
0.012&$4.18\times 10^{-18}$ &$4.28\times 10^{-18}$&0.6&$5.25\times 10^{1}$&$5.19\times 10^{1}$\\
0.013&$1.92\times 10^{-17}$ &$1.96\times 10^{-17}$&0.7&$1.26\times 10^{2}$&$1.25\times 10^{2}$\\
0.014&$7.59\times 10^{-17}$ &$7.77\times 10^{-17}$&0.8&$2.39\times 10^{2}$&$2.38\times 10^{2}$\\
0.015&$2.64\times 10^{-16}$ &$2.71\times 10^{-16}$&0.9&$3.88\times 10^{2}$&$3.87\times 10^{2}$\\
0.016&$8.28\times 10^{-16}$ &$8.48\times 10^{-16}$&1&$5.64\times 10^{2}$&$5.63\times 10^{2}$\\
0.018&$6.24\times 10^{-15}$ &$6.39\times 10^{-15}$&1.25&$1.06\times 10^{3}$&$1.06\times 10^{3}$\\
0.020&$3.55\times 10^{-14}$ &$3.63\times 10^{-14}$&1.5&$1.55\times 10^{3}$&$1.56\times 10^{3}$\\
0.025&$1.15\times 10^{-12}$ &$1.18\times 10^{-12}$&1.75&$1.96\times 10^{3}$&$1.98\times 10^{3}$\\
0.030&$1.64\times 10^{-11}$ &$1.68\times 10^{-11}$&2&$2.29\times 10^{3}$&$2.32\times 10^{3}$\\
0.040&$7.79\times 10^{-10}$ &$7.98\times 10^{-10}$&2.5&$2.72\times 10^{3}$&$2.76\times 10^{3}$\\
0.050&$1.21\times 10^{-8}$ &$1.24\times 10^{-8}$&3&$2.95\times 10^{3}$&$3.00\times 10^{3}$\\
0.060&$9.86\times 10^{-8}$ &$1.01\times 10^{-7}$&3.5&$3.05\times 10^{3}$&$3.11\times 10^{3}$\\
0.070&$5.25\times 10^{-7}$ &$5.38\times 10^{-7}$&4&$3.10\times 10^{3}$&$3.15\times 10^{3}$\\
0.080&$2.09\times 10^{-6}$ &$2.14\times 10^{-6}$&5&$3.11\times 10^{3}$&$3.16\times 10^{3}$\\
0.090&$6.73\times 10^{-6}$ &$6.90\times 10^{-6}$&6&$3.09\times 10^{3}$&$3.14\times 10^{3}$\\
0.10&$1.85\times 10^{-5}$  &$1.89\times 10^{-5}$&7&$3.08\times 10^{3}$&$3.13\times 10^{3}$\\
0.11&$4.47\times 10^{-5}$  &$4.58\times 10^{-5}$&8&$3.09\times 10^{3}$&$3.14\times 10^{3}$\\
0.12&$9.79\times 10^{-5}$  &$1.00\times 10^{-4}$&9&$3.14\times 10^{3}$&$3.18\times 10^{3}$\\
0.13&$1.98\times 10^{-4}$  &$2.03\times 10^{-4}$&10&$3.22\times 10^{3}$&$3.26\times 10^{3}$\\
\hline \hline
\end{tabular}\label{tab4}
\end{table*}

\begin{figure}[htbp]
\includegraphics[width=\columnwidth]{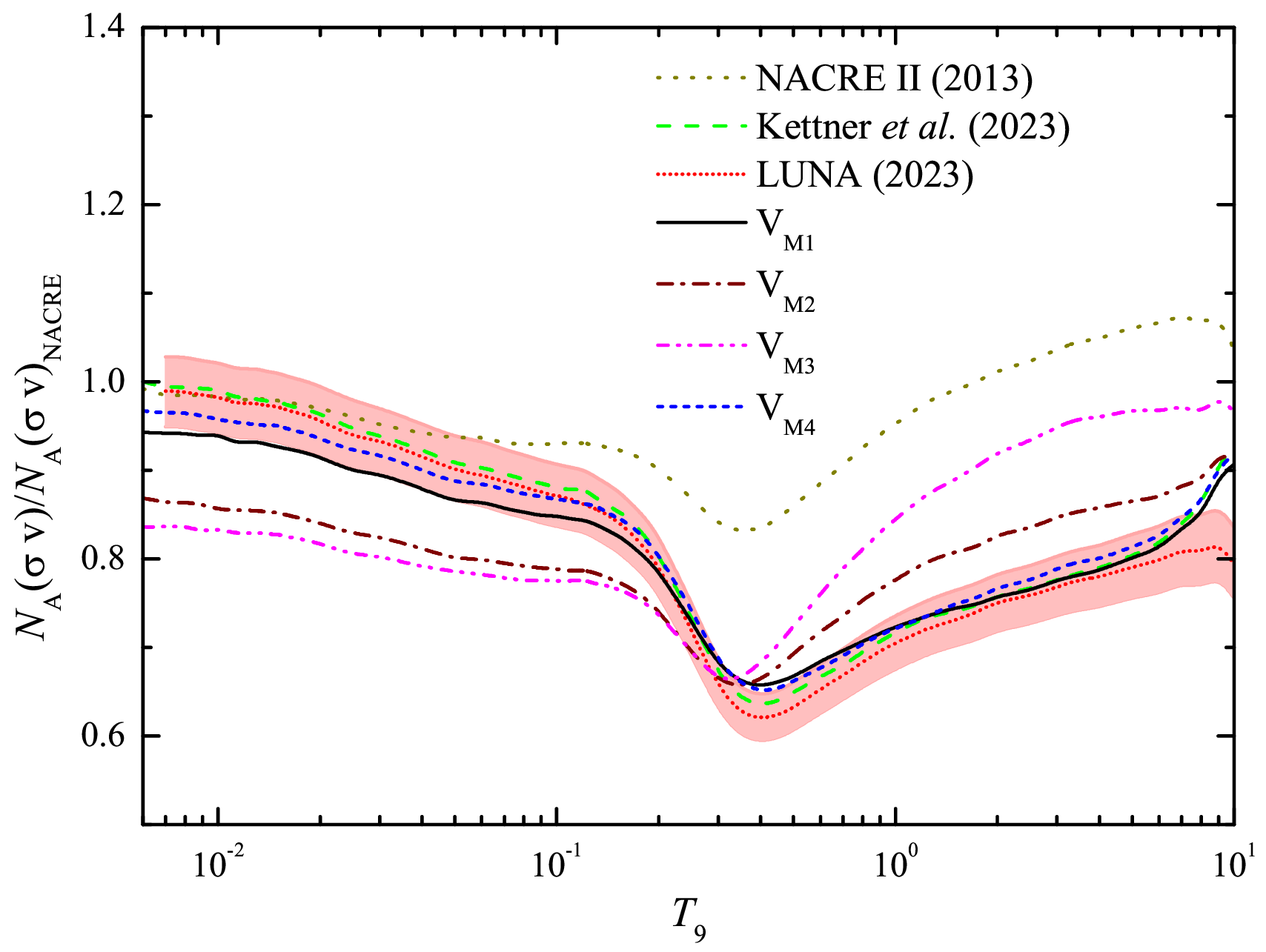}\caption{The
$^{12}$C(p,$\gamma)^{13}$N astrophysical reaction rates, normalized
to the NACRE rate \cite{nacre}, estimated within different potential
models, in comparison with the results of
Refs.~\cite{nacre2,kett23,luna23}.} \label{fig5}
\end{figure}

In Fig.~\ref{fig5} the calculated reaction rates within the
potential models $\textrm{V}_\textrm{M${\rm{1}}$}$,
$\textrm{V}_\textrm{M${\rm{2}}$}$, $\textrm{V}_\textrm{M${\rm{3}}$}$
and $\textrm{V}_\textrm{M${\rm{4}}$}$,  normalized to the NACRE rate
\cite{nacre}, are compared with the empirical results of Refs.
\cite{nacre2,kett23,luna23} in the temperature range from
$T_9$=0.006 to $T_9$=10, where $T_9$ is the temperature in the
10$^9$ K unit. The shaded area represents the uncertainty of the
empirical results of the LUNA Collaboration \cite{luna23}. As can be
seen in the figure, the models $\textrm{V}_\textrm{M${\rm{1}}$}$ and
$\textrm{V}_\textrm{M${\rm{4}}$}$ yield results, very consistent
with the results of the LUNA collaboration \cite{luna23} and those
from the R-matrix fit \cite{kett23}. They reproduce the both
absolute values and temperature dependence of the empirical results.
At the same time, the models $\textrm{V}_\textrm{M${\rm{2}}$}$ and
$\textrm{V}_\textrm{M${\rm{3}}$}$ do not give satisfactory results.
They underestimate the empirical results in the temperature interval
below the minimum point at $T_9$=0.4 ($T=4 \times 10^{-8}$ K) and
overestimate them in the temperature region above this point.

As we can remember from Table \ref{tab1}, the models
$\textrm{V}_\textrm{M${\rm{1}}$}$ and
$\textrm{V}_\textrm{M${\rm{4}}$}$  yield the ANC value 1.63
fm$^{-1/2}$ of the virtual decay $^{13}$ N(1/2$^-$)$\to ^{12}$C$+p$,
which turns out to be the most realistic. We also recall that these
models differ from each other only in the description of the width
of the $S$- wave $1/2^+$ resonance at the excitation energy
$E_x=2.365$ MeV. One can see from Fig.~\ref{fig5}, that the results
most close to the LUNA line belong to the model
$\textrm{V}_\textrm{M${\rm{4}}$}$, which yields the $1/2^+$
resonance width of 39 keV, while the model
$\textrm{V}_\textrm{M${\rm{1}}$}$  gives 36 keV. This means that the
reaction rates of the $^{12}$C$(p,\gamma)^{13}$N direct capture
process is extremely sensitive to the description of the $1/2^+$
resonance at $E_x=2.365$ MeV. The recommended theoretical value for
the width is slightly larger than the experimental widths of
$\Gamma=35.2\pm0.5$ keV~\cite{gyur23b} and $\Gamma=36.0\pm 1.9$
keV~\cite{rolfs74}. One can expect that this problem can be
clarified in further theoretical and experimental research studies.

\section{Conclusion}

The potential model was developed to study the astrophysical direct
nuclear capture process $^{12}$C($p,\gamma)^{13}$N. The Woods-Saxon
form of the phase-equivalent potentials for the $p-^{12}$C
interaction have been examined for their ability to describe the
experimental data on the astrophysical $S$ factor and the reaction
rates of the process. The parameters of the $p-^{12}$C potentials
are fitted to reproduce the bound state properties (binding energy
and the ANC), as well as the experimental phase shifts in the
scattering channels.

It was shown found, that the reaction rates of the process are very
sensitive to the description of the ANC of the $^{13}$N(1/2$^-$)
ground state and $S$- wave N($1/2^+$) resonance width at the
excitation energy $E_x=2.365$ MeV.  The potential model which
reproduce the empirical value of ANC, $C=1.63$ fm$^{-1/2}$ and the
width $\Gamma=39$ keV of the N$(1/2^+)$ resonance, yields a very
good description of the new experimental data for the astrophysical
$S$ factor and the empirical reaction rates of the LUNA
Collaboration. The calculated values of the astrophysical $S$ factor
at the solar Gamow energy are in good agreement with the result of
 $S(25~\rm{keV})=1.48 \pm 0.09$ keV b obtained by Kettner {\it et al.} \cite{kett23}
 using the R-matrix fit, but slightly lower than the result of the
LUNA Collaboration \cite{luna23,achar24} of $S(25~\rm{keV})=1.53 \pm
0.06$ keV b.

\section*{Acknowledgements}
The authors thank R.J. deBoer for sharing their results 
in a tabulated form.

 \end{document}